%% file: metastable_currop.tex
\documentclass[letterspace]{article}
\pdfoutput=1

\usepackage{jheppub,setspace}
\usepackage{url,comment}
\usepackage{times}
\usepackage{latexsym}
\usepackage{graphicx, graphics, hyperref, amsmath, amssymb, slashed, color, bbm,amsthm, array}
\usepackage[usenames,dvipsnames]{xcolor}
 \usepackage{subfigure}
\usepackage{mdwlist}
\usepackage{multirow}

\input{./sections/topmatter}

\input{./sections/titlepage}

\begin{document}

\setcounter{tocdepth}{3}
\maketitle

{\Large Highlights}\\
\begin{itemize}
\item Metastable activity is increasingly being observed in many cortical areas and in a variety of tasks.
\item Metastable activity is correlated with sensory and cognitive processes and may subserve state-dependent computations.
\item Metastable activity can emerge spontaneously in spiking networks with clustered architecture.
\item Metastable activity is a candidate neural substrate for coding both external events and internal representations.
\end{itemize}

\newpage

\input{./sections/Introduction}
\input{./sections/Main}

\input{./sections/Conclusions}

\input{./sections/Acknowledgments}

\section*{References and recommended reading}
Papers of particular interest, published within the period of review, have been highlighted as:

$\bullet$  of special interest

$\bullet\bullet$ of outstanding interest
\bigskip

$\bullet\bullet$ \cite{LitwinKumarDoiron2012} (Litwin-Kumar and Doiron, 2012)
This theoretical work demonstrated that a balanced spiking network with clustered architecture can generate slow metastable activity via internally generated fluctuations.

$\bullet\bullet$ \cite{mazzucato2015dynamics} (Mazzucato et al, 2015)
This study demonstrates that ongoing activity in rat gustatory cortex unfolds as a sequence of metastable states that is compatible with the dynamics of a clustered spiking network model. The model predicts key features of both ongoing and evoked activity, including a stimulus-induced reduction of multistability in single neurons. 

$\bullet\bullet$ \cite{schaub2015emergence} (Schaub et al, 2015)
This paper presents the link between metatable activity and the eigenvalue spectrum of the synaptic matrix in a clustered spiking network. When the model is augmented with a hierarchical architecture, transitions among metatable states at higher level in the hierarchy occur on slower timescales than transitions among the states at lower level of the hierarchy. 

$\bullet\bullet$ \cite{rich2016decoding} (Rich and Wallis, 2016)
In the OFC of monkeys, periods of deliberation during a choice task were associated with sequences of states representing the values (rewards) associated to the options being offered in the current choice. The longer the presence of one state compared to the others, the quicker the behavioral decision, regardless of the presumed difficulty of the choice according to the similarity in values. It is also shown that ensembles of OFC neurons, unlike single neurons, can represent the value of unchosen options.

$\bullet\bullet$ \cite{engel2016selective} (Engel et al, 2016)
In area V4 of behaving monkeys, ON and OFF metastable states are related to global changes in cortical state (associated with arousal) as well as local selective attention demands, with better performance when a stimulus change occurred during ON states compared to OFF states.

$\bullet\bullet$ \cite{maboudi2018uncovering} (Maboudi et al, 2018)
Spontaneous metastable states observed in CA1 during population burst events (when the animal is idling) reconstruct a map of place fields evoked during locomotion.

$\bullet\bullet$ \cite{taghia2018uncovering} (Taghia et al, 2018)
An HMM analysis of fMRI data in 122 human subjects performing a working memory task with multiple task conditions reveals the existence of metastable states preferentially associated to different task conditions. The hidden states are, in turn, associated to different patterns of functional connectivity across many brain areas.

$\bullet\bullet$ \cite{mazzucato2019expectation} (Mazzucato et al, 2019)
This paper shows that expectation induces faster coding of taste stimuli in rat gustatory cortex by modulating the timescale of ensemble metastable dynamics. A spiking network model explains the phenomenon in terms of decreased energy barriers between metastable configurations of the network?s activity. Shallower barriers cause faster transitions to stimulus-coding states identified with an HMM analysis, accelerating stimulus coding.

\bibliography{bib}
\bibliographystyle{JHEP}

\end{document}

%% file: sections/topmatter.tex
\numberwithin{equation}{section}

\def \eea{\end{eqnarray}}
\def \ee{\end{equation}}

\def\bc{\begin{center}}
\def\ec{\end{center}}

%% file: sections/titlepage.tex
\title{Cortical computations via metastable activity}

\author[1,2,*]{Giancarlo La Camera,}
\author[1,2]{Alfredo Fontanini,}
\author[3]{and Luca Mazzucato}

\affiliation[1]{Department of Neurobiology and Behavior and}
\affiliation[2]{Graduate Program in Neuroscience, State University of New York at Stony Brook, Stony Brook, NY 11794}
\affiliation[3]{Departments of Biology and Mathematics and Institute of Neuroscience, University of Oregon, Eugene, OR 94703}
\affiliation[*]{corresponding author}
  
\bigskip

\abstract{  
Metastable brain dynamics are characterized by abrupt, jump-like modulations so that the neural activity in single trials appears to unfold as a sequence of discrete, quasi-stationary `states.' Evidence that cortical neural activity unfolds as a sequence of metastable states is accumulating at fast pace. Metastable activity occurs both in response to an external stimulus and during ongoing, self-generated activity. These spontaneous metastable states are increasingly found to subserve internal representations that are not locked to external triggers, including states of deliberations, attention and expectation. Moreover, decoding stimuli or decisions via metastable states can be carried out trial-by-trial. Focusing on metastability will allow us to shift our perspective on neural coding from traditional concepts based on trial-averaging to models based on dynamic ensemble representations. 
Recent theoretical work has started to characterize the mechanistic origin and potential roles of  metastable representations. In this article we review recent findings on metastable activity, how it may arise in biologically realistic models, and its potential role for representing internal states as well as relevant task variables. 
\

\bigskip
\noindent Emails: $\textsf{giancarlo dot lacamera at stonybrook.edu}$, $\textsf{alfredo dot fontanini at stonybrook.edu}$,\\ $\textsf{luca dot mazzucato at uoregon dot edu}$
\noindent 
}


%% file: sections/Introduction.tex
\section{Introduction}

Brain circuits consist of large neural networks, where populations of neurons are recurrently coupled via synaptic connections. These circuits can be interpreted as dynamical systems with many coupled degrees of freedom and therefore capable of generating a wealth of dynamical behaviors on very diverse timescales \cite{bernacchia2011reservoir}. Those include transient relaxations towards a point or line attractor, oscillatory patterns, chaotic dynamics or metastable activity \cite{rabinovich2006dynamical,miller2016dynamical}. These dynamical behaviors have been implicated in important brain functions. Transient relaxation to stable neural activity (such as a point or line attractor) may subserve memory \cite{Amit1997b,seung1996brain} and perceptual decisions \cite{mante2013context,wang2002probabilistic}. Oscillatory dynamics may subserve respiration, locomotion, and other rhythmic forms of behavior \cite{wang2010neurophysiological,buzsaki2006rhythms}. Chaotic dynamics amplify random perturbations and, when successfully tamed by learning, can generate complex computations \cite{sussillo2009generating,laje2013robust}. Metastable dynamics, initially characterized in the presence of a sensory stimulus \cite{Abeles1995a,Jones2007}, might underlie a range of internal computations during cognitive tasks \cite{engel2016selective,PonceAlvarez2012,maboudi2018uncovering,rich2016decoding}.
Relaxations to an attractor, oscillations and chaotic dynamics typically describe neural activity as smoothly varying over time \cite{churchland2012neural,mazor2005transient}. In contrast, metastable dynamics are characterized by abrupt, jump-like modulations so that neural activity appears to unfold as a sequence of discrete, quasi-stationary ``states.'' Metastable activity is being found in an increasing number of brain structures of different species engaged in a variety of tasks, and is the focus of this opinion. The intuitive appeal of metastable activity is that it resonates with our intuition that our thoughts and actions proceed along a sequence of distinct episodes, as we scan alternatives and ponder potential options during everyday tasks. It is natural to think that such episodes are being represented in transient but well-defined neural patterns in our brains. Recent models have started to clarify how metastable activity could emerge spontaneously as a collective phenomenon via attractor dynamics in spiking networks \cite{LitwinKumarDoiron2012,DecoHugues2012,mazzucato2015dynamics,schaub2015emergence}. These models have proved powerful tools to investigate the circuit origin of metastable states and their potential role in sensory and cognitive processes. 
In this article we review recent progress in the analysis of metastable state sequences, focusing on issues of detection, modeling and interpretation.

%% file: sections/Main.tex

\begin{figure}[ht]
\begin{center}
\hspace*{-0.5cm}                                                           
\includegraphics[width=0.9\textwidth]{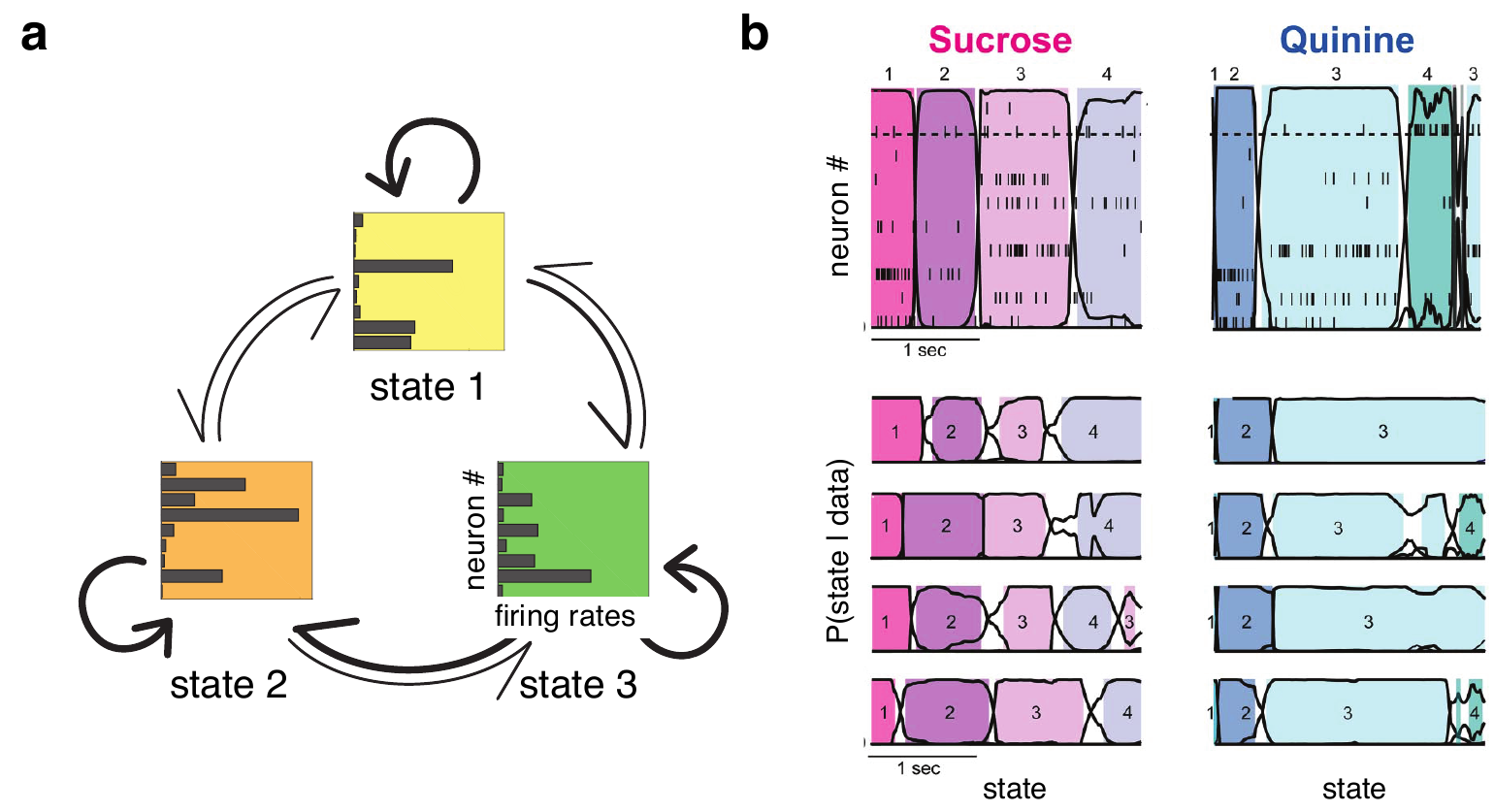}
\end{center}
\caption{{Observing metastability in neural populations \bf}. {\bf a:} Schematics illustration of a hidden Markov model (HMM) with three states. Hidden states (squares) are collections of firing rates across neurons; arrows indicate transition rates among the states (thicker arrows denote  larger transition rates). {\bf b:} HMM applied to electrophysiology recordings from the gustatory cortex of a behaving rat. The top panels show spike rasters from 10 simultaneously recorded neurons segmented via HMM analysis. Two trials are shown, one in which sucrose (left) and one in which quinine (right) was administered to the rat. HMM states were assigned to each bin of data when their posterior probability exceeded 0.8 (dashed horizontal line). The bottom panels show the state sequences in 4 different trials with either sucrose (left) or quinine (right) as a taste stimulus. Transitions may occur at variable times across trials, but the state sequences are reliable. Panel (b) modified from \cite{Jones2007}.
}
\label{Fig1}
\end{figure}

\section{Observing metastable activity}

Metastable states were first characterized in electrophysiological recordings from  the prefrontal cortex of monkeys performing a delayed localization task \cite{Abeles1995a,Seidemann1996,gat1993statistical}. Spike counts from simultaneously recorded neurons were analyzed with a hidden Markov model (HMM), a popular latent variable model for analyzing sequential data arising e.g. in time series analysis, speech recognition \cite{Rabiner1989}, DNA sequence analysis \cite{durbin1998biological}, behavioral analysis \cite{wiltschko2015mapping}, and many other problems \cite{zucchini2016hidden,dymarski2011hidden}. 
Under the most common application of HMM to neural data, population activity from simultaneously recorded neurons unfolds through a sequence of ``hidden'' states, each state being a set of ensemble firing rates (Fig. \ref{Fig1}). Stochastic transitions between states occur at random times according to an underlying Markov chain, and neurons discharge as Poisson processes with state-dependent firing rates, though the model can be extended to include refractory periods and other history-dependent factors \cite{Escola2011}. The number of hidden states may be selected using information criteria \cite{PonceAlvarez2012,sadacca2016behavioral,mazzucato2019expectation}, cross-validation \cite{engel2016selective}, or nonparametric Bayesian methods \cite{linderman2016bayesian,taghia2018uncovering}. Several measures have been used to assess the goodness-of-fit of inferred sequences, including variance explained \cite{engel2016selective}, comparison to shuffled datasets \cite{maboudi2018uncovering}, and comparison of single-neuron transitions to ensemble state transition \cite{mazzucato2015dynamics}. The strength of HMM is to allow a principled, unsupervised method for segmenting neural activity into a sequence of discrete metastable states, and its use has led to intriguing new observations in ensemble dynamics, as we discuss next.  

\subsection{Consistent features of metastable activity}

HMM and related procedures for detecting hidden states have established a number of consistent features across datasets, animals and tasks. Ensemble activity unfolds as a sequence of metastable states, each lasting from a few hundred ms to a second or more, with sharp transitions among the states \cite{Abeles1995a,Jones2007,PonceAlvarez2012,mazzucato2015dynamics} (Fig. \ref{Fig1}b). The transitions are typically one order of magnitude faster than the state durations, are close to their theoretically observable lower bound, and are not an artifact of the HMM \cite{Abeles1995a,Jones2007,PonceAlvarez2012}. When the start of the sequence is aligned to an external event such as stimulus onset, specific state sequences are observed (Fig. \ref{Fig1}b), raising the possibility that stimuli are coded by dynamic state sequences \cite{Abeles1995a,Jones2007}.
Among the most significant early results of applying HMM to the analysis of neural data were the demonstration that (i) the hidden states identified in response to a stimulus tend to recur during most of the recorded activity, not just in response to the stimulus, and (ii) pairwise correlations among simultaneously recorded neurons depend on an underlying global activity state, and not just on neural connectivity \cite{Abeles1995a,Seidemann1996}. These two observations are very relevant for current issues in neuroscience, including the origin and role of ongoing activity \cite{stringer2019spontaneous,orban2016neural}, the importance of the connectome \cite{swanson2016cajal,bargmann2012beyond}, and the meaning of correlations \cite{ostojic2009connectivity,doiron2016mechanics,rosenbaum2014correlated}. 
It's worth remarking that the ability of HMM analysis to segment the neural activity into distinct states in an unsupervised manner -- i.e., without knowledge of external triggers such as stimulus timings, -- has been crucial to obtaining the results reviewed above. It is also important to point out that the HMM analysis on ensemble data allows to decode the neural activity on a trial-by-trial basis, avoiding the necessity of averaging across trials. Most findings discussed in this review would go undetected if data were averaged across trials, because ensemble transitions occur at different times in each trial. This latter feature also allows the time-warping of state sequences, explaining why, for example, the method is so useful for speech recognition and for the analysis of neural activity during birdsongs \cite{florian2011hidden}. This could also explain part of the trial-to-trial variability observed in cortex after aligning spike trains across different trials to e.g.. stimulus onset \cite{Jones2007}.

\subsection{Metastability of ongoing cortical activity}
One central question in contemporary neuroscience is why ongoing cortical activity is so rich in structure and resembles so closely stimulus-evoked activity \cite{stringer2019spontaneous,orban2016neural,Arieli1996,Kenet2003,DecoJirsa2012,Luczak2009,Tsodyks1999,ringach2009spontaneous}. The origin of ongoing activity and the nature of its interaction with evoked activity can potentially reveal much about the way cortical networks are functionally organized and how this may support flexible coding (see \cite{stringer2019spontaneous,orban2016neural} for two recent proposals). However, quantifying precisely the similarities and differences between ongoing and evoked activity has proved difficult. A recent HMM analysis of ongoing spiking activity in the gustatory cortex of behaving rats has demonstrated that ongoing activity, similarly to evoked activity, unfolds as a sequence of metastable states \cite{mazzucato2015dynamics}. Transitions among states are characterized by a partial degree of coordination, so that ongoing activity is neither completely asynchronous nor completely synchronized. Ongoing and stimulus-evoked activity share most of the same states, although some of the states occur mostly during ongoing activity and others during evoked activity \cite{mazzucato2019expectation}. About $50\%$ of the ensemble neurons exhibit three or more different firing rates across states, i.e., they are ``multi-stable'' rather than bistable; and this fraction is substantially reduced by stimulation \cite{mazzucato2015dynamics}. These findings have all been captured by a spiking network model of metastability that might shed light into the link between ongoing and evoked activity (see section {Spiking network models of metastable activity\it}).

\subsection{Metastable sequences as a substrate for internal computations}

\begin{figure}[ht]
\vspace{-1cm}                                                        
\begin{center}
\hspace*{-0.4cm}                                                           
\includegraphics[width=0.9\textwidth]{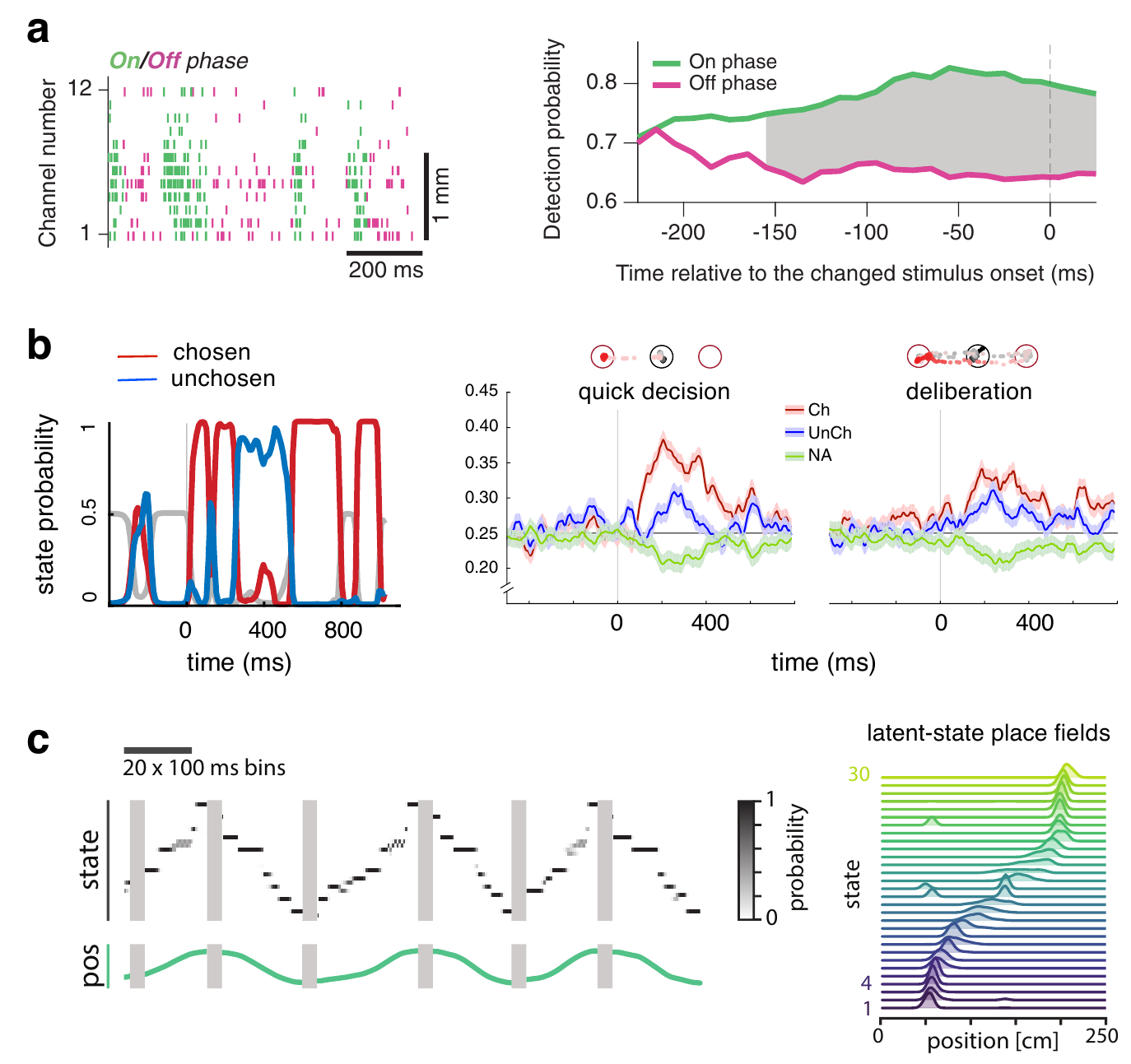}
\end{center}
\caption{{\bf Metastable activity and cognitive function}. {\bf a:} Left: Multiunit activity from monkey V4 during a selective attention task alternates between ON (green) and OFF (pink) HMM states. Right: Behavioral performance improves when the decision occurs during ON intervals compared to OFF intervals. {\bf b:} Left: Probability of being in a state representing the value of the chosen (red), unchosen (blue) and unavailable (gray) option from monkey OFC ensembles in a decision making task. Right panels: Comparison of time-course of state probability in quick vs. deliberative decisions as judged by the dynamics of eye movements (depicted at top). {\bf c:} Left: Dynamics of HMM states decoded from hippocampal neural activity (top) and position (bottom) of a rat running along a linear track (6 runs; only data during sharp wave ripples, comprising $2\%$ of the session, were used). Right: Mapping latent state probabilities to associated animal positions yields latent-state place fields describing the probability of each state for every position on the track. (a) panels modified from \cite{engel2016selective}; (b) panels modified from \cite{rich2016decoding}; (c) panels modified from \cite{maboudi2018uncovering}.
}
\label{Fig2}
\end{figure}

Do metastable states have precise and specific meanings as neural correlates of external or internally generated events? Visual \cite{Abeles1995a}, gustatory \cite{Jones2007,mazzucato2015dynamics} and vibro-tactile stimuli \cite{PonceAlvarez2012} seem represented by reliable sequences of HMM states in primate and rodent cortical areas in the context of different tasks. 
Even more remarkably, though, aspects of the metastable dynamics seem to subserve internal states of attention, expectation, and internal deliberations during decisions, and can be used to predict behavior. In monkeys performing a selective attention task,  the probability of detecting a change in the attended stimulus was significantly greater when occurring during an ``ON'' state (characterized by vigorous multi-unit activity in area V4) than during an ``OFF'' state (characterized by faint multi-unit activity) \cite{engel2016selective} (Fig. \ref{Fig2}a). Similarly, different hidden states were preferentially associated to different task conditions in a human working memory task, with occupancy rate in each state predicting better performance in the corresponding task \cite{taghia2018uncovering}. Moreover, changes in patterns of functional connectivity across many brain areas co-occurred more reliably with state transitions than with external triggers, suggesting that metastable states supported by flexible patterns of functional connectivity may reflect internal representations of task demands. 

A similar relationship between state occupancy rate and decisions has been found in monkeys performing a choice task between differentially rewarded stimuli \cite{rich2016decoding}. Ensemble activity in orbitofrontal cortex (OFC) mostly switched between two states representing the values of the stimuli currently being offered, with larger occupancy rate predicting the behavioral decision (Fig. \ref{Fig2}b, left). Intriguingly, slower decisions (reflecting longer internal deliberation) occurred when the states had similar occupancy rates and were found on trials of the same kind (thus reflecting subjective preferences rather than actual task difficulty; Fig. \ref{Fig2}b, right). These findings are reminiscent of those on dynamic changes of mind \cite{kiani2014dynamics} and complement earlier intriguing results on the possible meaning of state transitions, which may underlie the sudden realization that task rules have changed \cite{Durstewitz2010} or reflect the difficulty of a vibro-tactile discrimination \cite{PonceAlvarez2012}. 
In the rat hippocampus, HMM states were recently found to represent the position in a linear track and in an open field during a navigation task \cite{maboudi2018uncovering} (Fig. \ref{Fig2}c), an alternative view to decoding spatial maps based on single neurons place fields \cite{moser2017spatial}. The state sequences were observed in area CA1 during hippocampal sharp wave ripples when the animal was not exploring the track, and could be used to infer a spatial map of the environment without any reference to external locations \cite{maboudi2018uncovering}. 
In rat gustatory cortex, state sequences after stimulus delivery proceed faster when the delivery is expected compared to when it is not expected \cite{mazzucato2019expectation}, which may explain the faster decoding of expected stimuli found in this area \cite{samuelsen2012effects} (a spiking network model explanation of this phenomenon is discussed in the next section).
The examples discussed in this subsection make a convincing case that metastable sequences underlie a variety of important cognitive processes, and more examples will likely be found across different cognitive domains not yet explored with latent state models.

\section{Spiking network models of metastable activity}

In addition to statistical descriptions in terms of HMM and related latent state models, mechanistic accounts of metastable activity based on biologically plausible networks of spiking neurons have also been given. We shall mostly focus on efforts that have purposefully combined spiking network modeling with HMM \cite{mazzucato2015dynamics,mazzucato2019expectation,mazzucato2016stimuli,MillerKatz2010}, as we believe that such an approach is the most promising to understand the potential origin and function of metastable activity in neural circuits. Metastable activity naturally occurs when multiple hidden states are attractive fixed points of the neural dynamics which are either inherently unstable, or can be destabilized by internal noise or external perturbations \cite{miller2011stochastic,miller2016itinerancy,phillips2017cortical}. Such fixed points would attract the dynamic trajectories of the neural activity and force them to linger in their neighborhood for a finite amount of time. It is natural, therefore, to look for models wherein metastability is caused by the coexistence of multiple attractor states.

\subsection{Metastable activity in clustered networks of spiking neurons}

Among the class of attractor neural networks with multiple attractor states, spiking network models allow to build biologically plausible models capable of generating metastable state sequences. Recurrent networks with strong synaptic connections can produce highly temporally fluctuating activity \cite{sompolinsky1988chaos,rajan2010stimulus,ostojic2014two,huang2017once,harish2015asynchronous,mastrogiuseppe2017intrinsically,kadmon2015transition,wieland2015slow}, often the signature of chaotic activity. However, to endow a network of spiking neurons with endogenously generated metastable states, a specific partition of the network in subpopulations of neurons seems required 

\begin{figure}[ht]
\begin{center}
\hspace*{-0.2cm}                                                           
\includegraphics[width=0.9\textwidth]{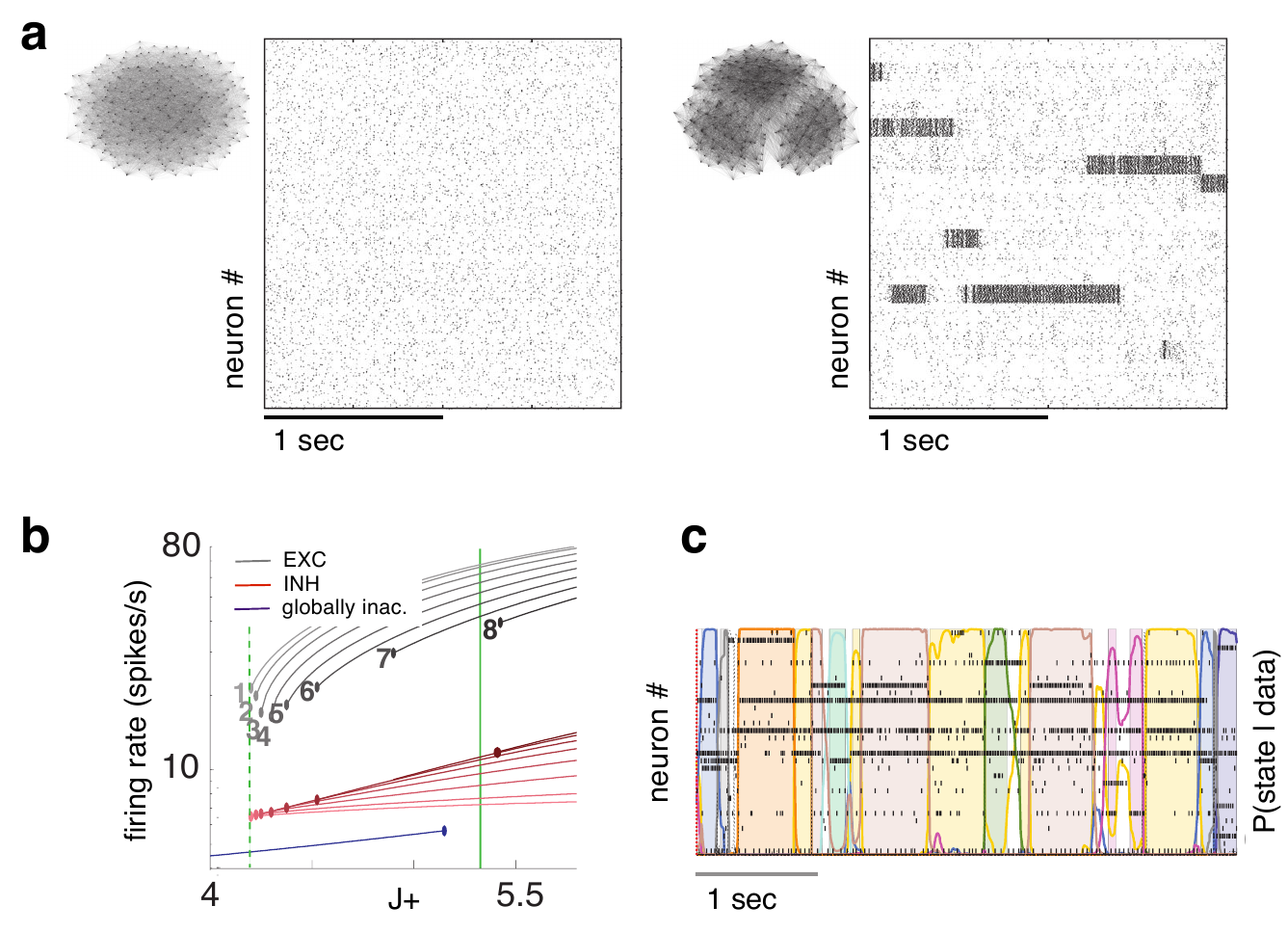}
\end{center}
\vspace*{-0.5cm}
\caption{{Spiking network model of metastable activity \bf}.  {\bf a:} Spike rasters of a spiking network model of excitatory (E) and inhibitory (I) integrate-and-fire neurons with two different architectures. Left: homogeneous network. Right: clustered network. The homogeneous network has uniform connectivity among the E neurons. In the clustered network, the E neurons are organized in clusters, so that synaptic connectivity is stronger inside each cluster than among clusters. The spiking activity in the clustered network is metastable. Insets: graphical description of the topology of each network. {\bf b:} The mean field theory of a clustered spiking network with 30 E clusters shows coexistence of several attractors for intra-cluster synaptic weight (ICW) $J_+$ beyond the critical value 4.2 (dashed vertical green line; $J_+$ is in units of the baseline synaptic weights outside clusters). The curves represent the firing rates of neurons in each cluster according to the number of active clusters (numbers from 1 to 8). Grey and purple curves: E neurons; red curves: I neurons. A state on the purple curve is ``globally inactive'' in the sense that no clusters are active and firing rates of E neurons remain low. {\bf c:} Raster plot from a typical simulation of the model in (b) with ICW corresponding to the full vertical green line. 30 E neurons (one for each separate cluster) are shown, together with the segmentation of the rasters in HMM states. Panel (a) adapted from \cite{LitwinKumarDoiron2012}; panel (b) adapted from \cite{mazzucato2016stimuli}; panel (c) adapted from \cite{mazzucato2015dynamics}.
}
\label{Fig3}
\end{figure}

\cite{LitwinKumarDoiron2012,DecoHugues2012,mazzucato2015dynamics,schaub2015emergence}. The clustered architecture has emerged as an effective way to generate metastability. The models studied so far comprise excitatory and inhibitory spiking neurons with (typically) the excitatory neurons organized in clusters, as seems the case in cortical circuits \cite{song2005highly,perin2011synaptic,kiani2015natural}. The common feature of these models is that the average synaptic strength (and/or connectivity) inside clusters is larger than between clusters (Fig. \ref{Fig3}a, right). A mean field analysis of these models \cite{Amit1997b,mazzucato2015dynamics,mazzucato2016stimuli} shows that multiple attractor states are possible depending on the average intra-cluster synaptic weight (ICW; Fig. \ref{Fig3}b). Above a critical ICW value (dashed green line in Fig. \ref{Fig3}b), multiple configurations of the network emerge wherein neurons in clusters tend to occupy several states with different values of mean firing rates (grey and red lines in Fig. \ref{Fig3}b), where the firing rates depend on the number of active clusters. In an infinite network, those configurations would be stable attractors; but in a finite network, or in the presence of sufficient external noise, those configurations become metastable. Note that if the ICW is large enough (e.g., full green line in Fig. \ref{Fig3}b), the configuration with no active clusters is unstable even in the infinite network, and at least one cluster must be active at any given time.
It is unclear whether suitable connectivity patterns exist such that metastable itinerant dynamics is possible also in a deterministic network of infinite size \cite{doiron2014balanced}. In the following, we assume that fluctuations originate in finite size effects in a deterministic network \cite{LitwinKumarDoiron2012,mazzucato2015dynamics}, and later consider alternative models (see subsection {\it Variations on the clustered architecture}).
In finite clustered networks above the critical ICW value, a combination of erratic spiking activity and recurrent inhibition can ignite an itinerant exploration of multiple configurations (as in Fig. \ref{Fig3}a, right), whose number and features depend on network parameters such as the ICW. When an HMM analysis is run on these networks, segmentation of the spontaneous ongoing activity in discrete states is found (Fig. \ref{Fig3}c). Such metastable activity is modified, but not removed, by an external stimulus. Since the firing rates depend on the number of active clusters, the neurons exhibit multiple firing rates across different HMM states, as found experimentally in rat gustatory cortex \cite{mazzucato2015dynamics}. Note that the neurons themselves can have graded activity, hence their multistability is an emergent property of the collective network's dynamics and not an intrinsic property of the units.

\subsection{Predictions of the clustered spiking network}

Clustered spiking networks capture many statistical and dynamical features of the data, including a stimulus-induced reduction of: (i) trial-to-trial variability \cite{LitwinKumarDoiron2012,DecoHugues2012,Churchland2010}, (ii) multistability of the firing rates across states \cite{mazzucato2015dynamics}, and (iii) neural dimensionality \cite{mazzucato2016stimuli}, i.e., the minimal number of effective dimensions (between 1 and the number of neurons) necessary to describe the ensemble dynamics \cite{rajan2010stimulus,williamson2016scaling,gao2017theory}. It is worth noticing that while the stimulus-driven reduction of trial-to-trial variability has been a catalyst for the development of the clustered network model \cite{LitwinKumarDoiron2012,DecoHugues2012}, the reduction of multistability and dimensionality were found to be natural emergent properties of the model which led to finding the corresponding properties in the data.
These predictions capture aspects of the data and thus help to corroborate the validity of the model. However, two other recent predictions link metastability directly with sensory and cognitive functions. The first is that visiting metastable states during ongoing activity will help to maintain learned stimulus responses and improve performance, suggesting yet another role for ongoing activity. This prediction emerged from an attempt to understand how neural clusters and metastable states can emerge during learning, and revealed the need to complement synaptic plasticity with homeostatic mechanisms \cite{ocker2015self,LitwinKumarDoiron2014,zenke2015diverse}. The second prediction has led to the proposal that a state of expectation could be understood as an acceleration of metastable dynamics \cite{mazzucato2019expectation}. When rats are trained to respond to taste stimuli, the identity of the stimuli can be decoded faster from the neural activity if stimulus delivery is expected compared to when stimuli are randomly, and passively, administered \cite{samuelsen2012effects}. In the model, the onset of a predictive cue modifies the network's landscape of metastable configurations so as to make state transitions more frequent. Faster state transitions, in turn, induce faster onset of stimulus-coding states after stimulus delivery, an effect uncovered by an HMM analysis of the model and confirmed in the data.

\subsection{Variations on the clustered architecture}

Variations of the clustered network architecture and its generated dynamics can explain related kinds of metastable activity such as perceptual bistability \cite{moreno2007noise,cao2016collective}, the emergence of specific slow oscillations known as ``up'' and ``down'' states \cite{setareh2017cortical}, or networks that can sustain bidirectional sequence propagation at slow and tunable speed \cite{setareh2018excitable}. These types of metastable dynamics are different from those reviewed earlier in subtle, but significant, ways. 
Clustered spiking networks have also been used to uncover subtle differences between mechanisms for decision making. Specifically, spiking networks with multiple attractors built for decision making \cite{wang2002probabilistic,wang2008decision} can work in two different modes, the ``ramping'' mode, characterized by gradual stimulus-driven firing rate increases, and the ``jumping'' mode \cite{miller2011stochastic,miller2016itinerancy}, wherein the stimulus ignites a sequence of metastable states. By leveraging a mechanism reminiscent of stochastic resonance, the jumping mode can outperform the ramping mode (underlying the so-called ``integration-to-bound'' models \cite{bogacz2006physics}) during perceptual decision-making in the face of sensory noise. 
Finally, we briefly mention that models with endogenously generated activity are not necessary to produce metastable activity. External fluctuations or the presence of firing rate adaptation both can create instability and lead to metastability in a finite network with multiple attractors \cite{Giugliano2008,russo2012cortical} (including ``up'' and ``down'' states at low firing rate, \cite{jercog2017up}), and sometimes both ingredients are required to reproduce the statistics of the empirical data \cite{shpiro2009balance,jercog2017up}. Also, it is possible to obtain stimulus-evoked quenching of variability in non-metastable networks, such as networks with chaotic attractors \cite{sompolinsky1988chaos,rajan2010stimulus,crisanti2018path} or the supralinear stabilized network (SSN) \cite{rubin2015stabilized,ahmadian2013analysis}. These models capture the reduction in variability via different mechanisms and make different predictions; for example, the SSN can capture the tuning-dependent modulation of variability observed in primary visual cortex \cite{hennequin2018dynamical}.

%% file: sections/Conclusions.tex
\section{Conclusions}
\label{discussion}

We have reviewed recent progress made in elucidating the nature and the role of metastable states in brain function. Thanks to the availability of ensemble data from behaving animals, the study of metastable activity is changing the way we think about neural coding for both external events and internal representations.
We have also reviewed a class of spiking network models that promise to provide a mechanistic understanding of the origin and function of metastability. These models have demonstrated the importance of a clustered architecture for the spontaneous generation of metastability, the coexistence of ongoing and evoked metastability, the benefits of metastable sequences to decision making, and the explanation of faster coding of expected events via the acceleration of metastable dynamics. Because of their biological plausibility, not only can these models shed light on the mechanisms, but they can also be powerful tools for analyzing metastable states, predict causal relationships to behavior, and help to design new experimental studies.
In the future, it will be important to establish a causal relationship between metastable states and behavior, by manipulating neural activity during behavioral tasks while recording large ensembles across multiple sites. It will also be important to establish the emergence of neural clusters and metastable dynamics though development or learning. Recurrent spiking network models, in conjunction with latent-state models such as HMM, are likely to play an important role in his endeavor.

%% file: sections/Acknowledgments.tex
\section*{Conflict of interest statement}

The authors declare no conflict of interest.

\section*{Acknowledgments}

The authors acknowledge support from the NSF (grant IIS-1161852 to GLC), the NIH (NIH/NIDCD R01DC015234 to AF and NIH/NIDCD K25DC013557 to LM) and the Swartz Foundation (Award 66438 to LM) during the development of some of the ideas and concepts discussed in this article.